\documentclass{SciPost}

\hypersetup{
 colorlinks,
 linkcolor={red!50!black},
 citecolor={blue!50!black},
 urlcolor={blue!80!black}
 }

\usepackage[bitstream-charter]{mathdesign}
\DeclareSymbolFont{usualmathcal}{OMS}{cmsy}{m}{n}
\DeclareSymbolFontAlphabet{\mathcal}{usualmathcal}

\fancypagestyle{SPstyle}{
\fancyhf{}
\lhead{\colorbox{scipostblue}{\bf \color{white} ~SciPost Physics }}
\rhead{{\bf \color{scipostdeepblue} ~Submission }}

\fancyfoot[C]{\textbf{\thepage}}
}

\usepackage{graphicx}
\usepackage{tabularx}
\usepackage{subcaption}
\usepackage{placeins}
\usepackage{multirow}
\usepackage{graphbox}
\usepackage{xcolor}
\usepackage{authblk}
\usepackage{tikz-feynman}
\tikzfeynmanset{compat=1.0.0}
\usepackage{hyperref}
\usepackage[numbers, sort&compress]{natbib}

\def\beq{\begin{equation}}
\def\eeq{\end{equation}}

\begin{document}

\numberwithin{equation}{section}

\pagestyle{SPstyle}

\begin{center}{\Large \textbf{\color{scipostdeepblue}{Triple-Higgs production: a window on the Higgs potential \\[2mm] and anomalous top-Higgs couplings}}}\end{center}

\begin{center}\textbf{
{John~M.~Campbell}\textsuperscript{1}, 
{Giuseppe~De~Laurentis}\textsuperscript{2}, 
{R.~Keith~Ellis}\textsuperscript{3}, 
{Michelle Gensmann}\textsuperscript{4,5},
{Ulrich~Haisch}\textsuperscript{4$\star$},
{Luca~Panizzi}\textsuperscript{6},
{Andreas~Papaefstathiou}\textsuperscript{7$\dagger$} and
{Gilberto~Tetlalmatzi-Xolocotzi}\textsuperscript{8}
}\end{center}

\begin{center}
{\bf 1 }{Fermilab, PO Box 500, Batavia IL 60510-5011, USA} \\
{\bf 2 }{CERN, Theoretical Physics Department, CH-1211 Geneva 23, Switzerland} \\
{\bf 3 }{Institute for Particle Physics Phenomenology, Durham University, Durham, DH1 3LE, UK} \\
{\bf 4 }{Max Planck Institute for Physics, Boltzmannstr. 8, 85748 Garching, Germany} \\
{\bf 5 }{Institute for Theoretical Physics, Karlsruhe Institute of Technology, \\ Wolfgang-Gaede-Str. 1, 76131 Karlsruhe, Germany} \\
{\bf 6 }{Dipartimento di Fisica, Università della Calabria, \\ I-87036 Arcavacata di Rende, Cosenza, Italy} \\
{\bf 7 }{Department of Physics, Kennesaw State University, \\ 830 Polytechnic Lane, Marietta, GA 30060, USA} \\
{\bf 8 }{Theoretische Physik 1, Center for Particle Physics Siegen (CPPS), \\ Universit{\"a}t Siegen, Walter-Flex-Str.~3, 57068 Siegen, Germany}
\\[\baselineskip]
$\star$ \href{mailto:haisch@mpp.mpg.de}{\small haisch@mpp.mpg.de} \qquad
$\dagger$ \href{mailto:apapaefs@kennesaw.edu}{\small apapaefs@kennesaw.edu} 
\end{center}
\begin{center}
 Preprint: FERMILAB-PUB-26-0464-T, LHCHWG-2026-017, MPP-2026-181
\end{center}

\section*{\color{scipostdeepblue}{Abstract}}
\textbf{\boldmath{%
We present a comprehensive study of triple-Higgs production within the~$\kappa$-framework as a probe of the Higgs potential and anomalous top-Higgs couplings. At leading order, we derive compact parametrisations of the inclusive $gg\to3h$ cross section in terms of the trilinear and quartic Higgs self-coupling modifiers~$\kappa_3$ and~$\kappa_4$, the top-quark Yukawa coupling modifier~$\kappa_t$, and the anomalous two-top-two-Higgs and two-top-three-Higgs coupling modifiers~$\kappa_{2t}$ and~$\kappa_{3t}$. These parametrisations allow for a fast and accurate interpretation of current and future LHC measurements. Applying them to LHC~Run~2 data and high-luminosity LHC projections, we quantify the impact of~$\kappa_t$,~$\kappa_{2t}$, and~$\kappa_{3t}$ on the constraints on~$\kappa_3$ and~$\kappa_4$, and derive the first direct constraint on~$\kappa_{3t}$ from existing LHC~data. We~further show that parameter points yielding nearly identical inclusive rates can give rise to different kinematic distributions, demonstrating the potential of differential measurements to resolve degeneracies and enhance the sensitivity of future triple-Higgs~searches.
}}

\vspace{\baselineskip}

\noindent\textcolor{white!90!black}{%
\fbox{\parbox{0.975\linewidth}{%
\textcolor{white!40!black}{\begin{tabular}{lr}%
 \begin{minipage}{0.6\textwidth}%
 {\small Copyright attribution to authors. \newline
 This work is a submission to SciPost Physics. \newline
 License information to appear upon publication. \newline
 Publication information to appear upon publication.}
 \end{minipage} & \begin{minipage}{0.4\textwidth}
 {\small Received Date \newline Accepted Date \newline Published Date}%
 \end{minipage}
\end{tabular}}
}}
}



\newpage

\vspace{10pt}
\noindent\rule{\textwidth}{1pt}
\tableofcontents
\noindent\rule{\textwidth}{1pt}
\vspace{10pt}

\allowdisplaybreaks

\section{\textbf{\boldmath{Introduction}}}
\label{sec:introduction}

The discovery of the Higgs boson in 2012~\cite{ATLAS:2012yve,CMS:2012qbp} completed the particle content of the Standard Model~(SM) and established the mechanism of electroweak~(EW) symmetry breaking. Since~then, a major objective of the LHC physics programme has been to determine the properties of the Higgs boson with increasing precision. While its couplings to EW gauge bosons and third-generation fermions are already constrained at the few- to ten-percent level, its self-interactions remain largely unexplored. Measuring the Higgs self-couplings is therefore one of the most important outstanding goals of collider physics, since they directly probe the shape of the Higgs potential and the dynamics of EW symmetry breaking.

In the SM, the Higgs potential, $V = \lambda \left (H^\dagger H - v^2/2 \right )^2$, is determined by a single quartic coupling $\lambda = m_h^2/(2v^2)$. After EW symmetry breaking, the Higgs doublet can be written in unitary gauge as $H = \big(0, (v+h)/\sqrt{2}\big)^T$, and the potential takes the form
\beq \label{eq:Vpot}
V = \frac{m_h^2}{2} \hspace{0.5mm} h^2 + \lambda_3 \hspace{0.25mm} v \hspace{0.25mm} h^3 + \lambda_4 \hspace{0.25mm} h^4 \,,
\eeq
which defines the trilinear and quartic Higgs self-couplings~$\lambda_3$ and~$\lambda_4$. The SM predicts unique values for these couplings,
\beq \label{eq:SM67}
\lambda_3^{\rm SM} = \lambda \,, \qquad
\lambda_4^{\rm SM} = \frac{\lambda}{4} \,,
\eeq
once the Higgs boson mass $m_h\simeq125\,\mathrm{GeV}$ and vacuum expectation value $v\simeq246\,\mathrm{GeV}$ are fixed. Any deviation from these predictions would provide evidence for physics beyond the SM, such as extended scalar sectors, composite Higgs models, or mechanisms for EW baryogenesis. Discussions of such beyond-the-SM~(BSM) scenarios can be found, for example, in~Refs.~\cite{DiMicco:2019ngk,Bahl:2022jnx,Durieux:2022hbu,Stylianou:2023tgg,Maura:2024zxz}.

At the LHC, double-Higgs production provides the primary direct probe of the trilinear Higgs self-coupling, while loop-induced effects in single-Higgs processes yield complementary constraints~\cite{Gorbahn:2016uoy,Degrassi:2016wml,Bizon:2016wgr,Maltoni:2017ims,Gorbahn:2019lwq,Haisch:2021hvy,Gao:2023bll,Haisch:2024nzv,Ghosh:2025fma,CMS:2026pnw}. These measurements are typically interpreted within the~$\kappa$-framework~\cite{LHCHiggsCrossSectionWorkingGroup:2012nn,LHCHiggsCrossSectionWorkingGroup:2013rie,LHCHiggsCrossSectionWorkingGroup:2016ypw}, which parametrises deviations from the SM in terms of coupling modifiers. For the Higgs trilinear self-coupling, one defines
\beq \label{eq:kappa3}
\kappa_3 = \frac{\lambda_3}{\lambda_3^{\rm SM}} \,.
\eeq
Using the full $\sqrt{s}=13\,\mathrm{TeV}$ LHC~Run~2 data set, the ATLAS and CMS collaborations obtain the combined $95\%$ confidence level~(CL) constraint~\cite{CMS:2026nuu}
\beq \label{eq:kappa3bound}
-0.71 < \kappa_3 < 6.1\;\;\;(\text{ATLAS}+\text{CMS}) \,,
\eeq
assuming all other Higgs couplings to take their SM values. With an integrated luminosity of $3\,\mathrm{ab}^{-1}$ per experiment, or $6\,\mathrm{ab}^{-1}$ in total for ATLAS and CMS, the high-luminosity LHC~(HL-LHC) is expected to improve this sensitivity~to
\beq \label{eq:kappa3boundfuture}
0.5 < \kappa_3 < 1.7\;\;\;(\text{HL-LHC})\,.
\eeq
This projection is based on the study presented in~Ref.~\cite{ATL-PHYS-PUB-2025-018}.

Constraining the quartic Higgs self-coupling is considerably more challenging. It is conventionally parametrised as
\beq \label{eq:kappa4}
\kappa_4 = \frac{\lambda_4}{\lambda_4^{\rm SM}} \,,
\eeq
and triple-Higgs production provides its only direct probe at hadron colliders. However, the tiny SM cross section of approximately $100\,\mathrm{ab}$ at $\sqrt{s}=14\,\mathrm{TeV}$, calculated at next-to-next-to-leading order~(NNLO) in QCD~\cite{deFlorian:2019app}, renders such measurements experimentally very challenging. Current ATLAS~\cite{ATLAS:2024xcs} and CMS~\cite{CMS:2026xkc} searches yield
\beq \label{eq:kappa4bounds}
-230 < \kappa_4 < 240 \;\;\;(\rm ATLAS) \,, \qquad
-177 < \kappa_4 < 185 \;\;\;(\rm CMS) \,,
\eeq
at the $95\%$~CL, assuming all other couplings are SM-like, in particular~$\kappa_3=1$. A recent HL-LHC study~\cite{Ding:2026qto} projects these bounds to improve to
\beq \label{eq:kappa4boundsHLLHC}
-81 < \kappa_4 < 89\;\;\;(\text{HL-LHC}) \,,
\eeq
for a combined integrated luminosity of $6\,\mathrm{ab}^{-1}$ from ATLAS and CMS. In addition to triple-Higgs production, double-Higgs production is also sensitive to~$\kappa_4$ through two-loop EW corrections~\cite{Bizon:2018syu,Borowka:2018pxx,Li:2024iio,Bizon:2024juq,Ding:2026qto}, although the resulting constraints are generally weaker. Quadruple-Higgs production likewise provides sensitivity to~$\kappa_4$, but again yields weaker direct constraints~\cite{Papaefstathiou:2026sfv}.

In this note, we present a comprehensive study of triple-Higgs production in gluon fusion within the~$\kappa$-framework. We derive compact parametrisations of the inclusive cross section in terms of the trilinear and quartic Higgs self-coupling modifiers~$\kappa_3$ and~$\kappa_4$, the top-quark Yukawa coupling modifier~$\kappa_t$, and the two-top-two-Higgs and two-top-three-Higgs coupling modifiers~$\kappa_{2t}$ and~$\kappa_{3t}$. Top-Higgs contact interactions are included because they can be sizeable in BSM scenarios in which the Higgs boson is realised non-linearly --- see, for instance,~Refs.~\cite{Gillioz:2012se,Contino:2013kra,Li:2019ghf,Agrawal:2019bpm}. The parametrisations provide a convenient framework for interpreting current and future LHC measurements. We use them to constrain~$\kappa_3$ and~$\kappa_4$, including the impact of correlated variations in~$\kappa_t$, and to study the effects of~$\kappa_{2t}$ and~$\kappa_{3t}$, deriving the first dedicated constraint on~$\kappa_{3t}$ from existing LHC data. Finally, we show how differential distributions provide additional sensitivity and resolve degeneracies among the~$\kappa$~parameters that the inclusive rate alone cannot. Perturbative-unitarity bounds on~$\kappa_{2t}$ and~$\kappa_{3t}$ are derived in~Appendix~\ref{app:unitarity}. A complementary study of triple-Higgs production in gluon fusion, carried out within the LHC Higgs Working Group, is presented in~Ref.~\cite{Bethani:2026}. That work has a different emphasis and also studies $t\bar{t}hh$ production. We encourage interested readers to consult both works in parallel.

\section{\textbf{\boldmath{Triple-Higgs production in the~$\kappa$-framework}}}
\label{sec:results}

\begin{figure}[t!]
\begin{center}
\includegraphics[width=0.85\textwidth]{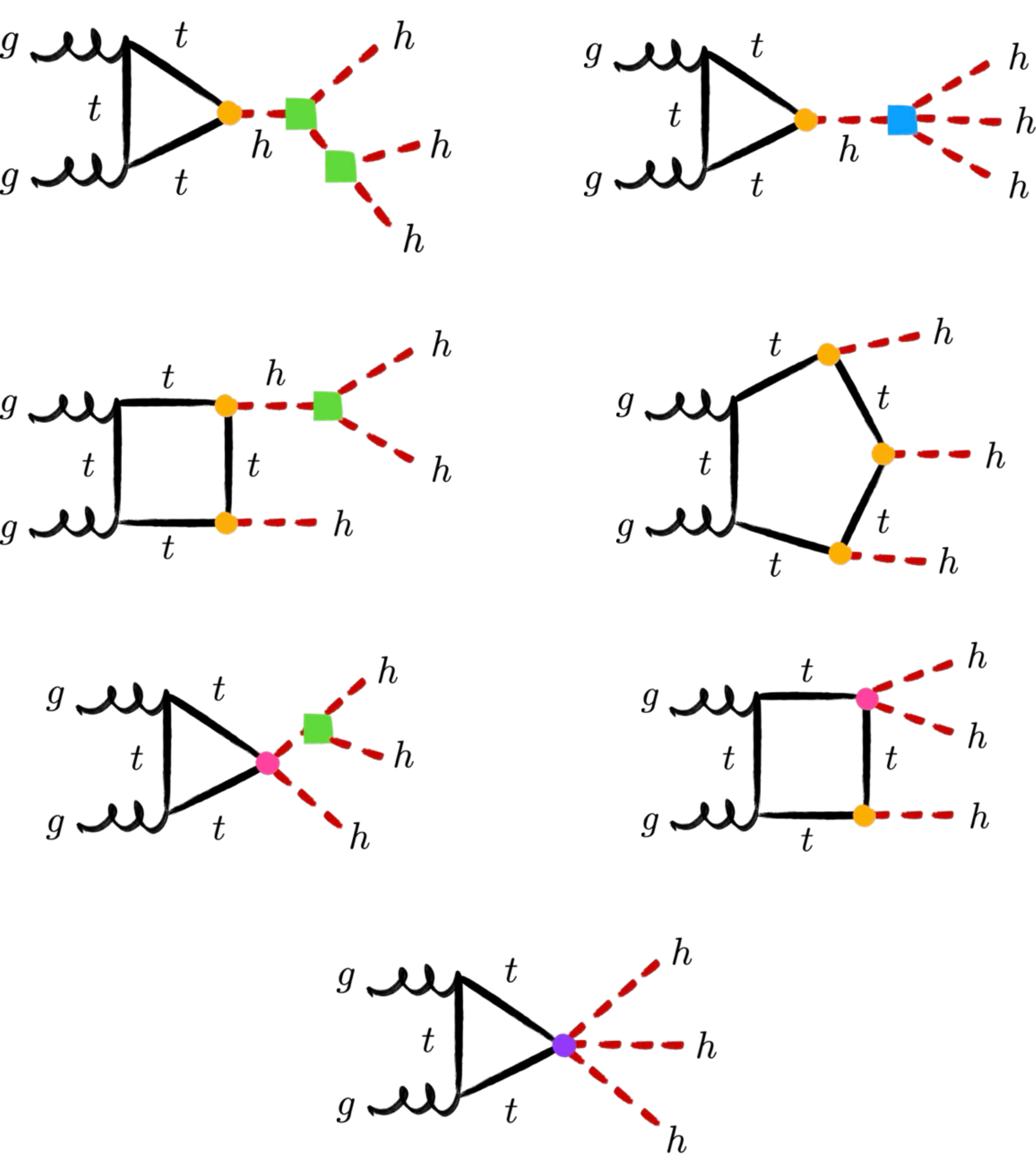} 
\end{center}
\vspace{2mm} 
\caption{\label{fig:diagrams} Representative one-loop Feynman diagrams contributing to triple-Higgs production via gluon fusion, showing the triangle, box, and pentagon topologies. Green and blue squares denote the trilinear and quartic Higgs self-couplings, respectively, while orange, magenta, and purple circles denote top-quark vertices coupling to one, two, and three Higgs bosons, respectively.}
\end{figure}

Triple-Higgs production at hadron colliders has been the subject of extensive theoretical investigation over the past two decades~\cite{Plehn:2005nk,Binoth:2006ym,Maltoni:2014eza,Papaefstathiou:2015paa,Chen:2015gva,Fuks:2015hna,deFlorian:2016sit,Kilian:2017nio,Fuks:2017zkg,Bizon:2018syu,deFlorian:2019app,Agrawal:2019bpm,Stylianou:2023tgg,Papaefstathiou:2023uum,Abouabid:2024gms,Bizon:2024juq,Dong:2025lkm,Haisch:2025pql,Campbell:2025llb,Fuks:2025gjv,Panizzi:2025sya,Ding:2026qto,Bethani:2026}. Within the SM, the gluon-fusion process $gg \to 3h$ is dominated by the one-loop topologies shown in the first and second rows of~Figure~\ref{fig:diagrams}. In the figure, green and blue squares denote insertions of the trilinear and quartic Higgs self-couplings, respectively, while orange circles represent the top-quark Yukawa coupling. The two diagrams in the third row illustrate representative BSM contributions involving a contact interaction between two top quarks and two Higgs bosons, denoted by a magenta circle, while the diagram in the fourth row represents a BSM contribution involving a contact interaction between two top quarks and three Higgs bosons, denoted by a purple circle. In~the~SM, bottom-quark loops modify the $gg \to 3h$ production cross section by only about $0.2\%$ at LHC energies~\cite{Campbell:2025llb}. We therefore neglect these contributions in the numerical calculations presented below.

The relevant terms in the $\kappa$-framework Lagrangian, which parametrises the interaction vertices highlighted in Figure~\ref{fig:diagrams}, can be written as
\beq \label{eq:Lkappa}
{\cal L}_{\kappa} \supset -\kappa_3 \hspace{0.25mm} v \lambda h^3 -\kappa_4 \hspace{0.25mm} \frac{\lambda}{4} \hspace{0.5mm} h^4 -\kappa_t \hspace{0.25mm} \frac{y_t}{\sqrt{2}} \hspace{0.5mm} \bar t t h -\kappa_{2t} \hspace{0.25mm} \frac{y_t}{\sqrt{2} \hspace{0.25mm} v} \hspace{0.5mm} \bar t t h^2 -\kappa_{3t} \hspace{0.25mm} \frac{y_t}{\sqrt{2} \hspace{0.25mm} v^2} \hspace{0.5mm} \bar t t h^3 \,,
\eeq
where~$\kappa_3$ and~$\kappa_4$ were introduced in~Eqs.~(\ref{eq:kappa3}) and (\ref{eq:kappa4}), respectively, and $y_t = \sqrt{2} \hspace{0.125mm} m_t/v$, with $m_t$ denoting the top-quark mass. The parameter~$\kappa_t$ thus parametrises modifications of the top-quark Yukawa coupling, while~$\kappa_{2t}$ and~$\kappa_{3t}$ parametrise the contact interactions between two top quarks with two and three Higgs bosons, respectively. The SM is recovered from~Eq.~(\ref{eq:Lkappa}) for~$\kappa_3 = \kappa_4 = \kappa_t = 1$ and~$\kappa_{2t} = \kappa_{3t} = 0$. We note that different normalisations and nomenclatures for the terms in~Eq.~(\ref{eq:Lkappa}) are used in the literature. See, for instance,~Refs.~\cite{Papaefstathiou:2023uum,Abouabid:2024gms,Buchalla:2018yce,Brivio:2025sib}. The normalisation adopted in~Eq.~(\ref{eq:Lkappa}) resembles the convention commonly used in the Higgs effective field theory~(HEFT) framework~\cite{Buchalla:2018yce,Brivio:2025sib}.

To determine the dependence of the inclusive $gg \to 3h$ production cross section on~$\kappa_3$,~$\kappa_4$, $\kappa_t$,~$\kappa_{2t}$, and~$\kappa_{3t}$, we employ the leading-order~(LO) matrix elements calculated in~Ref.~\cite{Campbell:2025llb} and implemented in the {\tt MCFM}~\cite{Campbell:1999ah,Campbell:2011bn,Campbell:2015qma,Campbell:2019dru} parton-level event generator. We take $m_h = 125\,\mathrm{GeV}$, $m_t = 173\,\mathrm{GeV}$, and $v = 246.22\,\mathrm{GeV}$ as input parameters and use the {\tt PDF4LHC21} parton distribution functions~(PDFs)~\cite{PDF4LHCWorkingGroup:2022cjn}. The renormalisation and factorisation scales are chosen dynamically as $\mu_R = \mu_F = m_{3h}/2$, where $m_{3h}$ denotes the invariant mass of the three-Higgs system. We also generate loop-induced LO parton-level samples with {\tt MadGraph5\_aMC@NLO}~\cite{Alwall:2014hca}, using the same input parameters, PDFs, and scale choices. The two approaches yield identical results to within about $1\%$, providing an important validation of our independent numerical calculations of the cross sections. 

\subsection{\texorpdfstring{\textbf{\boldmath{Inclusive cross section}}}{Inclusive cross section}}
\label{sec:inclusive}

At centre-of-mass~(COM) energies of $\sqrt{s}=13\,\mathrm{TeV}$, $\sqrt{s}=13.6\,\mathrm{TeV}$, and $\sqrt{s}=14\,\mathrm{TeV}$, we parametrise the LO cross section as
\beq \label{eq:BSMLO}
\sigma_{\mathrm{LO}}^{\sqrt{s}} = \big(\sigma_{\mathrm{LO}}^{\sqrt{s}}\big)_{\rm SM} \, \sum_{n,m} \, c_{nm}^{\sqrt{s}} \; \big(\Delta\kappa_3\big)^n \, \big(\Delta\kappa_4\big)^m \,,
\eeq
where $\Delta\kappa_i=\kappa_i-1$ for $i=3,4$. The corresponding LO SM cross sections are
\beq \label{eq:SMLO}
\big( \sigma_{\mathrm{LO}}^{13\,\mathrm{TeV}} \big)_{\mathrm{SM}} = 42.0\,\mathrm{ab} \,, \qquad
\big( \sigma_{\mathrm{LO}}^{13.6\,\mathrm{TeV}} \big)_{\mathrm{SM}} = 47.3\,\mathrm{ab} \,, \qquad
\big( \sigma_{\mathrm{LO}}^{14\,\mathrm{TeV}} \big)_{\mathrm{SM}} = 51.0\,\mathrm{ab} \,.
\eeq

For $\sqrt{s} = 13\,\mathrm{TeV}$, the coefficients $c_{nm}^{\sqrt{s}}$ appearing in~Eq.~(\ref{eq:BSMLO}) are
\begin{align} 
c_{00}^{13\,\mathrm{TeV}} & = 
0.106\hspace{0.25mm}\kappa_t^2
-0.682\hspace{0.25mm}\kappa_t^3
+1.93\hspace{0.25mm}\kappa_t^4
-3.29\hspace{0.25mm}\kappa_t^5
+2.94\hspace{0.25mm}\kappa_t^6 \notag \\[1mm]
& \phantom{=} +\kappa_{2t}
\left(
1.61\hspace{0.25mm}\kappa_t
-7.83\hspace{0.25mm}\kappa_t^2
+15.6\hspace{0.25mm}\kappa_t^3
-14.8\hspace{0.25mm}\kappa_t^4
\right) 
+\kappa_{2t}^2
\left(
6.98
-22.4\hspace{0.25mm}\kappa_t
+37.1\hspace{0.25mm}\kappa_t^2
\right) \notag \\[1mm]
& \phantom{=} +\kappa_{3t}
\left(
3.75\hspace{0.25mm}\kappa_t
-11.5\hspace{0.25mm}\kappa_t^2
+0.464\hspace{0.25mm}\kappa_t^3
+37.7\hspace{0.25mm}\kappa_{2t}
-60.2\hspace{0.25mm}\kappa_{2t}\hspace{0.125mm}\kappa_t
\right) 
+67.8\hspace{0.25mm}\kappa_{3t}^2 \,, \label{eq:c00_13} \\[2mm]
c_{10}^{13\,\mathrm{TeV}} & = 
0.257\hspace{0.25mm}\kappa_t^2
-1.52\hspace{0.25mm}\kappa_t^3
+3.68\hspace{0.25mm}\kappa_t^4
-3.29\hspace{0.25mm}\kappa_t^5 \notag \\[1mm]
& \phantom{=} +\kappa_{2t}
\left(
3.49\hspace{0.25mm}\kappa_t
-13.4\hspace{0.25mm}\kappa_t^2
+15.6\hspace{0.25mm}\kappa_t^3
\right) 
+\kappa_{2t}^2
\left(
14.0
-22.4\hspace{0.25mm}\kappa_t
\right) \notag \\[1mm]
& \phantom{=} +\kappa_{3t}
\left(
4.05\hspace{0.25mm}\kappa_t
-11.5\hspace{0.25mm}\kappa_t^2
+37.7\hspace{0.25mm}\kappa_{2t}
\right) \,, \label{eq:c10_13} \\[2mm]
c_{20}^{13\,\mathrm{TeV}} & = 
0.287\hspace{0.25mm}\kappa_t^2
-1.25\hspace{0.25mm}\kappa_t^3
+1.84\hspace{0.25mm}\kappa_t^4 \notag \\[1mm]
& \phantom{=} +\kappa_{2t}
\left(
2.82\hspace{0.25mm}\kappa_t
-6.71\hspace{0.25mm}\kappa_t^2
\right) 
+6.98\hspace{0.25mm}\kappa_{2t}^2 
+2.02\hspace{0.25mm}\kappa_{3t}\hspace{0.125mm}\kappa_t \,, \label{eq:c20_13} \\[2mm]
c_{30}^{13\,\mathrm{TeV}} & = 
0.159\hspace{0.25mm}\kappa_t^2
-0.418\hspace{0.25mm}\kappa_t^3
+0.940\hspace{0.25mm}\kappa_{2t}\hspace{0.125mm}\kappa_t \,, \label{eq:c30_13} \\[2mm]
c_{40}^{13\,\mathrm{TeV}} & = 
0.0397\hspace{0.25mm}\kappa_t^2 \,, \label{eq:c40_13} \\[2mm]
c_{01}^{13\,\mathrm{TeV}} & = 
0.0830\hspace{0.25mm}\kappa_t^2
-0.265\hspace{0.25mm}\kappa_t^3
+0.0897\hspace{0.25mm}\kappa_t^4
+\kappa_{2t}
\left(
0.674\hspace{0.25mm}\kappa_t
-1.12\hspace{0.25mm}\kappa_t^2
\right)
+1.73\hspace{0.25mm}\kappa_{3t}\hspace{0.125mm}\kappa_t \,, \label{eq:c01_13} \\[2mm]
c_{11}^{13\,\mathrm{TeV}} & = 
0.0980\hspace{0.25mm}\kappa_t^2
-0.265\hspace{0.25mm}\kappa_t^3
+0.674\hspace{0.25mm}\kappa_{2t}\hspace{0.125mm}\kappa_t \,, \label{eq:c11_13} \\[2mm]
c_{21}^{13\,\mathrm{TeV}} & = 
0.0490\hspace{0.25mm}\kappa_t^2 \,, \label{eq:c21_13} \\[2mm]
c_{02}^{13\,\mathrm{TeV}} & = 
0.0170\hspace{0.25mm}\kappa_t^2 \,. \label{eq:c02_13}
\end{align}

At $\sqrt{s} = 13.6\,\mathrm{TeV}$, the corresponding coefficients read 
\begin{align} 
c_{00}^{13.6\,\mathrm{TeV}} & = 
0.105\hspace{0.25mm}\kappa_t^2
-0.677\hspace{0.25mm}\kappa_t^3
+1.92\hspace{0.25mm}\kappa_t^4
-3.27\hspace{0.25mm}\kappa_t^5
+2.92\hspace{0.25mm}\kappa_t^6 \notag \\[1mm]
& \phantom{=} +\kappa_{2t}
\left(
1.61\hspace{0.25mm}\kappa_t
-7.79\hspace{0.25mm}\kappa_t^2
+15.5\hspace{0.25mm}\kappa_t^3
-14.8\hspace{0.25mm}\kappa_t^4
\right) 
+\kappa_{2t}^2
\left(
6.99
-22.4\hspace{0.25mm}\kappa_t
+37.3\hspace{0.25mm}\kappa_t^2
\right) \notag \\[1mm]
& \phantom{=} +\kappa_{3t}
\left(
3.76\hspace{0.25mm}\kappa_t
-11.6\hspace{0.25mm}\kappa_t^2
+0.348\hspace{0.25mm}\kappa_t^3
+38.0\hspace{0.25mm}\kappa_{2t}
-60.7\hspace{0.25mm}\kappa_{2t}\hspace{0.125mm}\kappa_t
\right) 
+69.3\hspace{0.25mm}\kappa_{3t}^2 \,, \label{eq:c00_136} \\[2mm]
c_{10}^{13.6\,\mathrm{TeV}} & = 
0.255\hspace{0.25mm}\kappa_t^2
-1.50\hspace{0.25mm}\kappa_t^3
+3.66\hspace{0.25mm}\kappa_t^4
-3.27\hspace{0.25mm}\kappa_t^5 \notag \\[1mm]
& \phantom{=} +\kappa_{2t}
\left(
3.48\hspace{0.25mm}\kappa_t
-13.4\hspace{0.25mm}\kappa_t^2
+15.5\hspace{0.25mm}\kappa_t^3
\right) 
+\kappa_{2t}^2
\left(
14.0
-22.4\hspace{0.25mm}\kappa_t
\right) \notag \\[1mm]
& \phantom{=} +\kappa_{3t}
\left(
4.04\hspace{0.25mm}\kappa_t
-11.6\hspace{0.25mm}\kappa_t^2
+38.0\hspace{0.25mm}\kappa_{2t}
\right) \,, \label{eq:c10_136} \\[2mm]
c_{20}^{13.6\,\mathrm{TeV}} & = 
0.285\hspace{0.25mm}\kappa_t^2
-1.24\hspace{0.25mm}\kappa_t^3
+1.83\hspace{0.25mm}\kappa_t^4 \notag \\[1mm]
& \phantom{=} +\kappa_{2t}
\left(
2.80\hspace{0.25mm}\kappa_t
-6.68\hspace{0.25mm}\kappa_t^2
\right) 
+6.99\hspace{0.25mm}\kappa_{2t}^2 
+2.02\hspace{0.25mm}\kappa_{3t}\hspace{0.125mm}\kappa_t \,, \label{eq:c20_136} \\[2mm]
c_{30}^{13.6\,\mathrm{TeV}} & = 
0.157\hspace{0.25mm}\kappa_t^2
-0.414\hspace{0.25mm}\kappa_t^3
+0.935\hspace{0.25mm}\kappa_{2t}\hspace{0.125mm}\kappa_t \,, \label{eq:c30_136} \\[2mm]
c_{40}^{13.6\,\mathrm{TeV}} & = 
0.0394\hspace{0.25mm}\kappa_t^2 \,, \label{eq:c40_136} \\[2mm]
c_{01}^{13.6\,\mathrm{TeV}} & = 
0.0825\hspace{0.25mm}\kappa_t^2
-0.263\hspace{0.25mm}\kappa_t^3
+0.0878\hspace{0.25mm}\kappa_t^4
+\kappa_{2t}
\left(
0.673\hspace{0.25mm}\kappa_t
-1.12\hspace{0.25mm}\kappa_t^2
\right)
+1.74\hspace{0.25mm}\kappa_{3t}\hspace{0.125mm}\kappa_t \,, \label{eq:c01_136} \\[2mm]
c_{11}^{13.6\,\mathrm{TeV}} & = 
0.0973\hspace{0.25mm}\kappa_t^2
-0.263\hspace{0.25mm}\kappa_t^3
+0.673\hspace{0.25mm}\kappa_{2t}\hspace{0.125mm}\kappa_t \,, \label{eq:c11_136} \\[2mm]
c_{21}^{13.6\,\mathrm{TeV}} & = 
0.0487\hspace{0.25mm}\kappa_t^2 \,, \label{eq:c21_136} \\[2mm]
c_{02}^{13.6\,\mathrm{TeV}} & = 
0.0169\hspace{0.25mm}\kappa_t^2 \,. \label{eq:c02_136}
\end{align}

Finally, for $\sqrt{s} = 14\,\mathrm{TeV}$, we obtain
\begin{align} 
c_{00}^{14\,\mathrm{TeV}} & = 
0.104\hspace{0.25mm}\kappa_t^2
-0.674\hspace{0.25mm}\kappa_t^3
+1.91\hspace{0.25mm}\kappa_t^4
-3.25\hspace{0.25mm}\kappa_t^5
+2.91\hspace{0.25mm}\kappa_t^6 \notag \\[1mm]
& \phantom{=} +\kappa_{2t}
\left(
1.60\hspace{0.25mm}\kappa_t
-7.77\hspace{0.25mm}\kappa_t^2
+15.5\hspace{0.25mm}\kappa_t^3
-14.7\hspace{0.25mm}\kappa_t^4
\right) 
+\kappa_{2t}^2
\left(
6.99
-22.4\hspace{0.25mm}\kappa_t
+37.4\hspace{0.25mm}\kappa_t^2
\right) \notag \\[1mm]
& \phantom{=} +\kappa_{3t}
\left(
3.76\hspace{0.25mm}\kappa_t
-11.6\hspace{0.25mm}\kappa_t^2
+0.274\hspace{0.25mm}\kappa_t^3
+38.2\hspace{0.25mm}\kappa_{2t}
-60.9\hspace{0.25mm}\kappa_{2t}\hspace{0.125mm}\kappa_t
\right) 
+70.3\hspace{0.25mm}\kappa_{3t}^2 \,, \label{eq:c00_14} \\[2mm]
c_{10}^{14\,\mathrm{TeV}} & = 
0.253\hspace{0.25mm}\kappa_t^2
-1.50\hspace{0.25mm}\kappa_t^3
+3.64\hspace{0.25mm}\kappa_t^4
-3.25\hspace{0.25mm}\kappa_t^5 \notag \\[1mm]
& \phantom{=} +\kappa_{2t}
\left(
3.47\hspace{0.25mm}\kappa_t
-13.3\hspace{0.25mm}\kappa_t^2
+15.5\hspace{0.25mm}\kappa_t^3
\right) 
+\kappa_{2t}^2
\left(
14.0
-22.4\hspace{0.25mm}\kappa_t
\right) \notag \\[1mm]
& \phantom{=} +\kappa_{3t}
\left(
4.04\hspace{0.25mm}\kappa_t
-11.6\hspace{0.25mm}\kappa_t^2
+38.2\hspace{0.25mm}\kappa_{2t}
\right) \,, \label{eq:c10_14} \\[2mm]
c_{20}^{14\,\mathrm{TeV}} & = 
0.283\hspace{0.25mm}\kappa_t^2
-1.23\hspace{0.25mm}\kappa_t^3
+1.82\hspace{0.25mm}\kappa_t^4 \notag \\[1mm]
& \phantom{=} +\kappa_{2t}
\left(
2.79\hspace{0.25mm}\kappa_t
-6.66\hspace{0.25mm}\kappa_t^2
\right) 
+6.99\hspace{0.25mm}\kappa_{2t}^2 
+2.02\hspace{0.25mm}\kappa_{3t}\hspace{0.125mm}\kappa_t \,, \label{eq:c20_14} \\[2mm]
c_{30}^{14\,\mathrm{TeV}} & = 
0.156\hspace{0.25mm}\kappa_t^2
-0.411\hspace{0.25mm}\kappa_t^3
+0.931\hspace{0.25mm}\kappa_{2t}\hspace{0.125mm}\kappa_t \,, \label{eq:c30_14} \\[2mm]
c_{40}^{14\,\mathrm{TeV}} & = 
0.0391\hspace{0.25mm}\kappa_t^2 \,, \label{eq:c40_14} \\[2mm]
c_{01}^{14\,\mathrm{TeV}} & = 
0.0822\hspace{0.25mm}\kappa_t^2
-0.262\hspace{0.25mm}\kappa_t^3
+0.0866\hspace{0.25mm}\kappa_t^4
+\kappa_{2t}
\left(
0.673\hspace{0.25mm}\kappa_t
-1.12\hspace{0.25mm}\kappa_t^2
\right)
+1.75\hspace{0.25mm}\kappa_{3t}\hspace{0.125mm}\kappa_t \,, \label{eq:c01_14} \\[2mm]
c_{11}^{14\,\mathrm{TeV}} & = 
0.0969\hspace{0.25mm}\kappa_t^2
-0.262\hspace{0.25mm}\kappa_t^3
+0.673\hspace{0.25mm}\kappa_{2t}\hspace{0.125mm}\kappa_t \,, \label{eq:c11_14} \\[2mm]
c_{21}^{14\,\mathrm{TeV}} & = 
0.0485\hspace{0.25mm}\kappa_t^2 \,, \label{eq:c21_14} \\[2mm]
c_{02}^{14\,\mathrm{TeV}} & = 
0.0169\hspace{0.25mm}\kappa_t^2 \,. \label{eq:c02_14}
\end{align}
The coefficients $c_{nm}^{\sqrt{s}}$ depend only mildly on the COM energy over the range considered, and since they are normalised to the SM cross section, the choice of PDFs is likewise expected to have a negligible impact on them. Their relative MC uncertainties, including an estimate of the residual dependence on the number of \texttt{VEGAS}~\cite{Lepage:1977sw} calls, are at the level of a few per mille, so that the third digit of some coefficients is uncertain by one or two units. The exception is the small $\kappa_{3t}\hspace{0.25mm}\kappa_t^3$ term in $c_{00}^{\sqrt{s}}$, which arises from large cancellations and carries an uncertainty of $2\%\hspace{0.25mm}$--$\hspace{0.25mm}4\%$, so that only its first two digits are significant. The parametrisations are therefore also provided in machine-readable form as ancillary files with the arXiv submission of this note, at full numerical precision and together with the MC uncertainties of all coefficients and their correlations. Higher-order QCD corrections to the coefficients $c_{nm}^{\sqrt{s}}$ are currently unknown, although they could in principle be estimated in the infinite-top-quark-mass limit. Within~the~SM, these corrections increase the inclusive cross section by a $K$-factor of approximately $1.7$ at NNLO in QCD~\cite{deFlorian:2019app}. Similar parametrisations of the triple-Higgs gluon-fusion cross section at the LHC have also been presented in~Refs.~\cite{Papaefstathiou:2023uum,Abouabid:2024gms,Haisch:2025pql,Campbell:2025llb,Bethani:2026}.

\begin{figure}[t!]
\begin{center}
\includegraphics[width=0.45\textwidth]{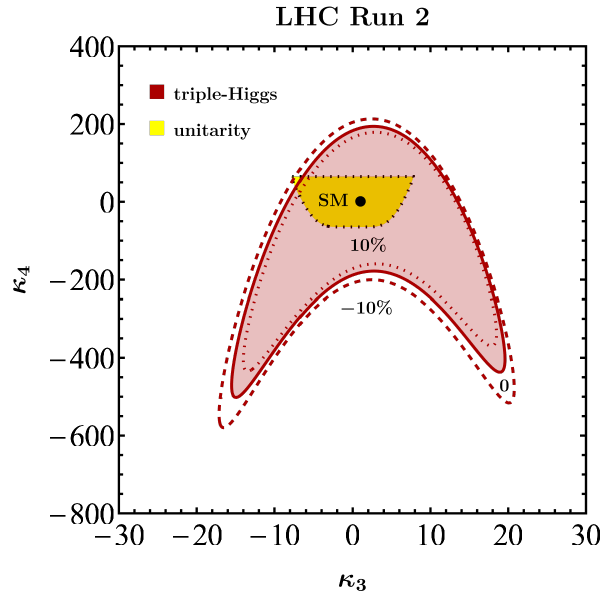} \quad 
\includegraphics[width=0.45\textwidth]{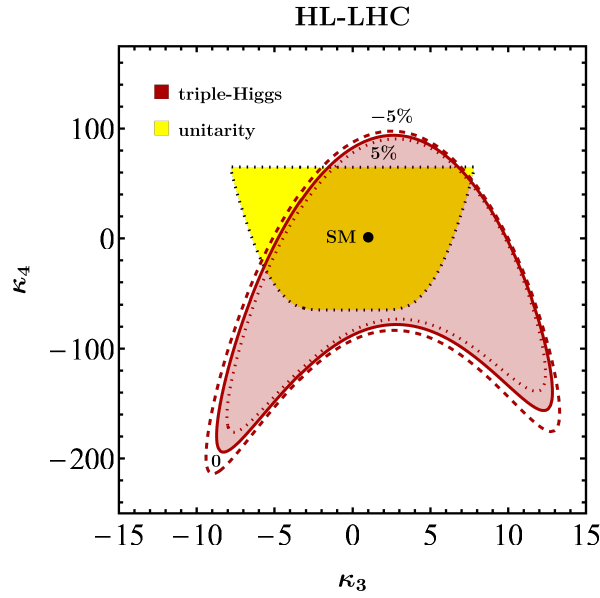}
\end{center}
\vspace{-4mm} 
\caption{\label{fig:planes} Constraints in the $\kappa_3\hspace{0.25mm}$--$\hspace{0.25mm}\kappa_4$ plane from LHC~Run~2~(left) and the HL-LHC~(right). The red contours show the allowed $95\%$~CL regions from triple-Higgs production in gluon fusion. Solid red contours correspond to $\kappa_t=1$, while dashed~(dotted) contours correspond to $\Delta\kappa_t=-10\%$ ($\Delta\kappa_t=10\%$) at LHC~Run~2 and $\Delta\kappa_t=-5\%$ ($\Delta\kappa_t=5\%$) at the HL-LHC, with $\Delta\kappa_t=\kappa_t-1$. We set $\kappa_{2t}=\kappa_{3t}=0$. The SM prediction is indicated by the black points. The yellow regions bounded by black dotted contours show the perturbative-unitarity bounds from tree-level $hh\to hh$ scattering.}
\end{figure}

The two panels in~Figure~\ref{fig:planes} illustrate the sensitivity to the Higgs self-couplings~$\kappa_3$ and~$\kappa_4$ based on the full LHC~Run~2 data set and HL-LHC projections, with an integrated luminosity of~$3\,\mathrm{ab}^{-1}$ per experiment. The red contours indicate the $95\%$~CL constraints derived from triple-Higgs production in gluon fusion using the parametrisations in~Eqs.~(\ref{eq:c00_13})--(\ref{eq:c02_13}) and~Eqs.~(\ref{eq:c00_14})--(\ref{eq:c02_14}), with the normalisation to the corresponding LO SM cross section given in~Eq.~(\ref{eq:SMLO}). For LHC~Run~2, we impose the upper limit $\mu_{3h}^{\text{LHC~Run~2}} < 588$~\cite{CMS:2026xkc}, while for the HL-LHC we assume the projected upper limit $\mu_{3h}^{\text{HL-LHC}} < 125$, consistent with~\cite{ATL-PHYS-PUB-2025-003}. Throughout this analysis, we set~$\kappa_{2t}=\kappa_{3t}=0$, and the black points indicate the SM prediction. The~left panel shows that shifts of $\Delta\kappa_t=\pm10\%$, where $\Delta\kappa_t=\kappa_t-1$, can visibly affect the resulting constraints. Such shifts remain allowed at $95\%$~CL by LHC~Run~2 data~\cite{ATLAS:2022vkf,CMS:2022dwd}. This behaviour was already pointed out in~Ref.~\cite{Panizzi:2025sya} and originates from the strong power-law dependence of the coefficients $c_{nm}^{\sqrt{s}}$ appearing in~Eqs.~(\ref{eq:c00_13})--(\ref{eq:c02_14}) on~$\kappa_t$. By the end of the HL-LHC, the top-quark Yukawa coupling is expected to be measured with a precision of $\Delta\kappa_t=\pm5\%$ at the~$95\%$~CL~\cite{ATL-PHYS-PUB-2025-018}. Consequently, its impact on the $gg\to3h$ constraints in the~$\kappa_3$--$\kappa_4$ plane will be reduced, as illustrated in the right panel. Finally, we note that the projected HL-LHC constraints from $gg\to3h$ are comparable to those imposed by perturbative unitarity~\cite{Liu:2018peg,Stylianou:2023tgg,DiLuzio:2017tfn}, shown as the yellow region enclosed by the black dotted contour in both~panels.

\begin{figure}[t!]
\begin{center}
\includegraphics[width=0.45\textwidth]{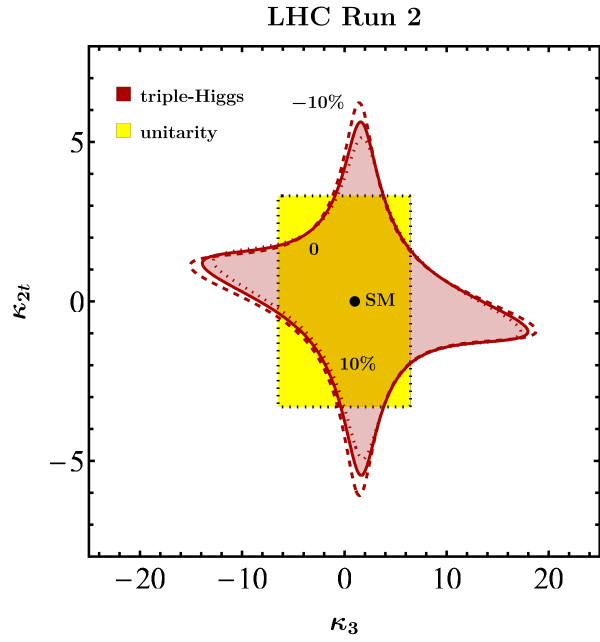} \quad 
\includegraphics[width=0.45\textwidth]{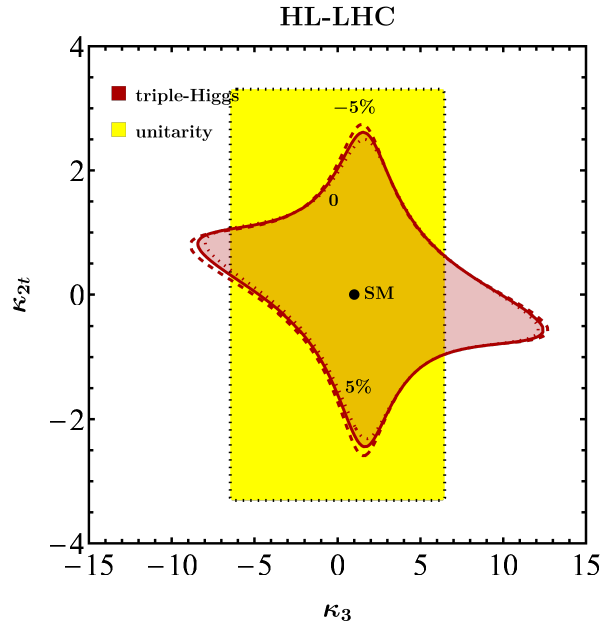}
\end{center}
\vspace{-4mm} 
\caption{\label{fig:planes2} Same as Figure~\ref{fig:planes}, but in the $\kappa_3\hspace{0.25mm}$--$\hspace{0.25mm}\kappa_{2t}$ plane at LHC~Run~2 (left) and the HL-LHC (right). We fix $\kappa_4=1$ and $\kappa_{3t}=0$. Solid red contours correspond to $\kappa_t=1$, while dashed and dotted contours correspond to the variations specified in the caption of~Figure~\ref{fig:planes}. The yellow regions bounded by black dotted contours show the perturbative-unitarity bounds from tree-level $hh \to hh$ and $t \bar t \to hh$ scattering.}
\end{figure}

\begin{figure}[t!]
\begin{center}
\includegraphics[width=0.45\textwidth]{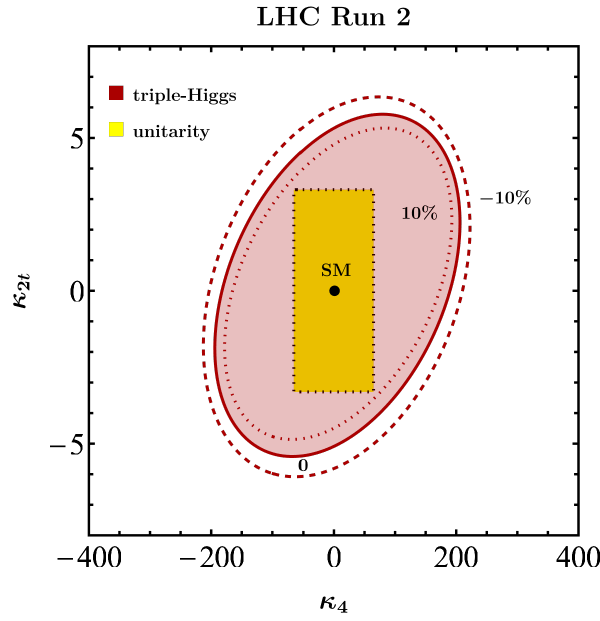} \quad 
\includegraphics[width=0.45\textwidth]{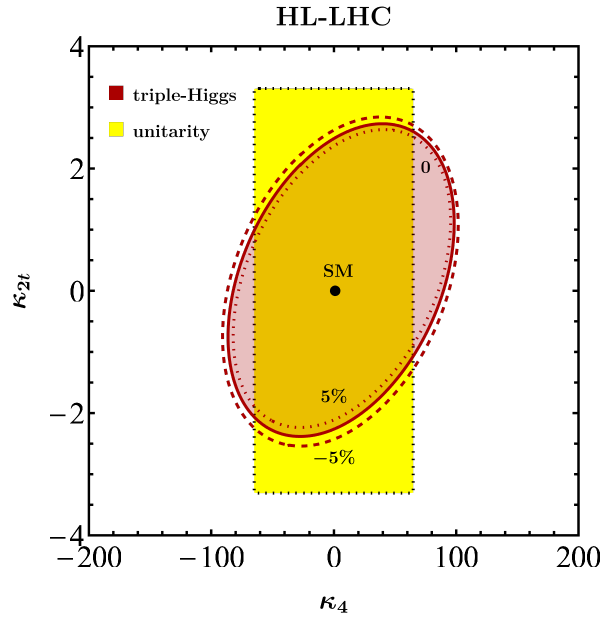}
\end{center}
\vspace{-4mm} 
\caption{\label{fig:planes3} Same as~Figure~\ref{fig:planes2}, but in the~$\kappa_4\hspace{0.25mm}$--$\hspace{0.25mm}\kappa_{2t}$ plane at LHC~Run~2~(left) and the HL-LHC~(right). We fix $\kappa_3=1$ and $\kappa_{3t}=0$. Solid red contours correspond to $\kappa_t=1$, while dashed and dotted contours correspond to the variations specified in the caption of~Figure~\ref{fig:planes}. The yellow regions bounded by black dotted contours show the perturbative-unitarity bounds from tree-level $hh \to hh$ and $t \bar t \to hh$ scattering.}
\end{figure}

\begin{figure}[t!]
\begin{center}
\includegraphics[width=0.45\textwidth]{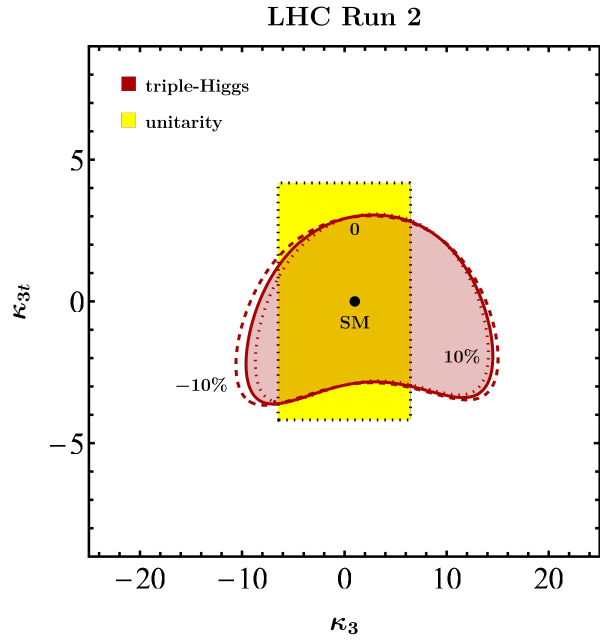} \quad 
\includegraphics[width=0.45\textwidth]{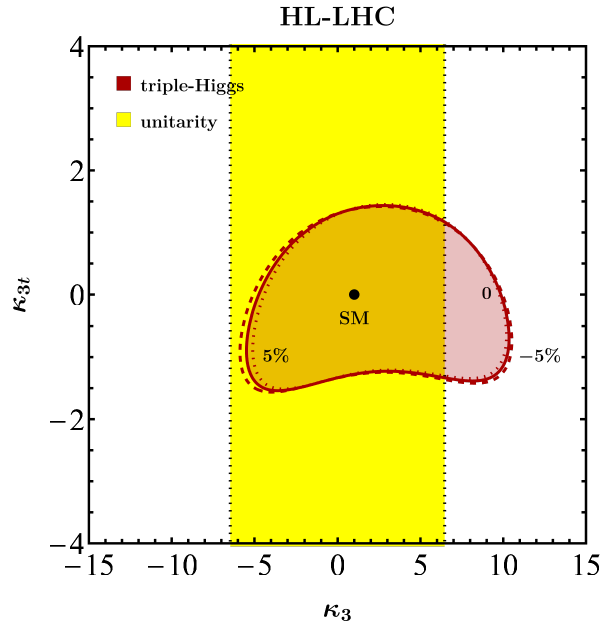}
\end{center}
\vspace{-4mm} 
\caption{\label{fig:planes4} Same as Figure~\ref{fig:planes}, but in the $\kappa_3\hspace{0.25mm}$--$\hspace{0.25mm}\kappa_{3t}$ plane at LHC~Run~2 (left) and the HL-LHC (right). We fix $\kappa_4=1$ and $\kappa_{2t}=0$. Solid red contours correspond to $\kappa_t=1$, while dashed and dotted contours correspond to the variations specified in the caption of~Figure~\ref{fig:planes}. The yellow regions bounded by black dotted contours show the perturbative-unitarity bounds from tree-level $hh \to hh$ and $t \bar t \to 3h$ scattering.}
\end{figure}

\begin{figure}[t!]
\begin{center}
\includegraphics[width=0.45\textwidth]{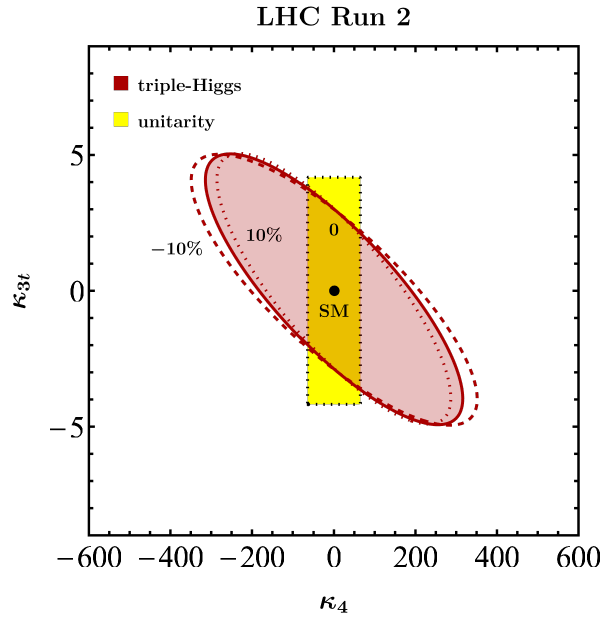} \quad 
\includegraphics[width=0.45\textwidth]{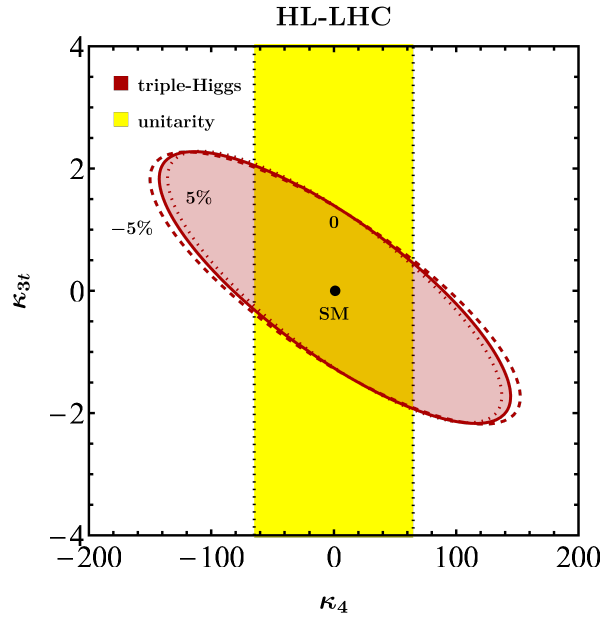}
\end{center}
\vspace{-4mm} 
\caption{\label{fig:planes5} Same as Figure~\ref{fig:planes3}, but in the $\kappa_4\hspace{0.25mm}$--$\hspace{0.25mm}\kappa_{3t}$ plane at LHC~Run~2 (left) and the HL-LHC (right). We fix $\kappa_3=1$ and $\kappa_{2t}=0$. Solid red contours correspond to $\kappa_t=1$, while dashed and dotted contours correspond to the variations specified in the caption of~Figure~\ref{fig:planes}. The yellow regions bounded by black dotted contours show the perturbative-unitarity bounds from tree-level $hh \to hh$ and $t \bar t \to 3h$ scattering.}
\end{figure}

The panels in~Figures~\ref{fig:planes2} and~\ref{fig:planes3} show the sensitivity in the~$\kappa_3\hspace{0.25mm}$--$\hspace{0.25mm}\kappa_{2t}$ and~$\kappa_4\hspace{0.25mm}$--$\hspace{0.25mm}\kappa_{2t}$ planes, respectively. The red contours represent the $95\%$~CL constraints from triple-Higgs production in gluon fusion, obtained following the procedure described above. Apart from $\kappa_t$, which varies as indicated by the contour labels, all parameters not shown on the axes are fixed to their SM values, and the black points indicate the SM prediction. As in~Figure~\ref{fig:planes}, the left panels show that variations of the top-quark Yukawa coupling by $\Delta\kappa_t=\pm10\%$ can have a visible impact on the resulting constraints, whereas the smaller variations $\Delta\kappa_t=\pm5\%$ considered in the right panels lead to only minor changes. For~$\kappa_{2t}$, triple-Higgs production at LHC~Run~2~yields
\beq \label{eq:k2t_hhh_LHCRunII}
-5.1 < \kappa_{2t} < 5.3 \,, 
\eeq
while the HL-LHC is expected to tighten this range to
\beq \label{eq:k2t_hhh_HLLHC}
-2.3 < \kappa_{2t} < 2.5 \,, 
\eeq
at $95\%$~CL. These limits are obtained by fixing~$\kappa_3=\kappa_4=\kappa_t=1$ and~$\kappa_{3t}=0$. Under the same assumptions, we note that the $95\%$~CL constraint on the double-Higgs production signal strength obtained at LHC~Run~2, $\mu_{2h}^{\text{LHC~Run~2}}<2.5$~\cite{CMS:2026nuu}, already implies
\beq \label{eq:k2t_hh_LHCRunII}
-0.19 < \kappa_{2t} < 0.72 \,,
\eeq
where the quoted bound is obtained using the expressions given in~Refs.~\cite{Buchalla:2018yce,Brivio:2025sib}. Note~that the bound in~Eq.~(\ref{eq:k2t_hhh_LHCRunII}) is weaker than the perturbative-unitarity bound on~$\kappa_{2t}$ derived in~Eq.~(\ref{eq:boundsnume}), while the bounds in~Eqs.~(\ref{eq:k2t_hhh_HLLHC}) and~(\ref{eq:k2t_hh_LHCRunII}) are stronger. The bound given in~Eq.~(\ref{eq:boundsnume}) is also indicated in~Figures~\ref{fig:planes2} and~\ref{fig:planes3} by the yellow regions enclosed by the black dotted contour. Triple-Higgs production is therefore not expected to provide a stronger constraint on~$\kappa_{2t}$ than double-Higgs production in a single-parameter analysis. It can, however, offer complementary information in multi-parameter fits, in particular when these involve~$\kappa_4$, which is only weakly constrained by double-Higgs~\cite{Bizon:2018syu,Borowka:2018pxx,Li:2024iio,Bizon:2024juq,Ding:2026qto} and quadruple-Higgs~\cite{Papaefstathiou:2026sfv} production.

Figures~\ref{fig:planes4} and~\ref{fig:planes5} illustrate the sensitivity in the~$\kappa_3\hspace{0.25mm}$--$\hspace{0.25mm}\kappa_{3t}$ and~$\kappa_4\hspace{0.25mm}$--$\hspace{0.25mm}\kappa_{3t}$ planes, respectively. The red contours show the $95\%$~CL constraints from triple-Higgs production in gluon fusion, obtained following the procedure described above. Except for~$\kappa_t$, which is varied as indicated by the contour labels, all parameters not shown on the axes are fixed to their SM values, and the black points indicate the SM prediction. As in~Figure~\ref{fig:planes}, the left panels demonstrate that variations of the top-quark Yukawa coupling by $\Delta\kappa_t=\pm10\%$ can visibly affect the resulting constraints, whereas the smaller variations of $\Delta\kappa_t=\pm5\%$ considered in the right panels lead to only modest changes. For the coupling modifier~$\kappa_{3t}$, triple-Higgs production at LHC~Run~2~provides
\beq \label{eq:k3t_hhh_LHCRunII}
-2.9 < \kappa_{3t} < 3.0 \,,
\eeq
while the HL-LHC is expected to improve this range to
\beq \label{eq:k3t_hhh_HLLHC}
-1.3 < \kappa_{3t} < 1.4 \,,
\eeq
at $95\%$~CL. The bounds given in~Eqs.~(\ref{eq:k3t_hhh_LHCRunII}) and~(\ref{eq:k3t_hhh_HLLHC}) are obtained by fixing $\kappa_3=\kappa_4=\kappa_t=1$ and~$\kappa_{2t}=0$. Both bounds are stronger than the perturbative-unitarity bound on~$\kappa_{3t}$ derived in~Eq.~(\ref{eq:boundsnume}), which is also shown in~Figures~\ref{fig:planes4} and~\ref{fig:planes5} by the yellow regions enclosed by the black dotted contour. To~our knowledge, the limit in~Eq.~(\ref{eq:k3t_hhh_LHCRunII}) constitutes the first constraint on the top-Higgs contact interaction parametrised by~$\kappa_{3t}$ derived from existing LHC~data. HL-LHC projections comparable to~Eq.~(\ref{eq:k3t_hhh_HLLHC}) have also been obtained in~Refs.~\cite{Papaefstathiou:2023uum,Abouabid:2024gms}. As~for~$\kappa_4$, $gg\to3h$ production therefore provides a particularly sensitive probe of~$\kappa_{3t}$, since this coupling modifier contributes to the process already at first order in perturbation theory.

The pronounced sensitivity to~$\kappa_{3t}$ also has important implications for the size of the triple-Higgs signal that can be accommodated within current and projected constraints. For instance, if $\kappa_3$ and $\kappa_4$ are varied freely, while $\kappa_t=1$ and $\kappa_{2t}=\kappa_{3t}=0$, the constraint from~Eq.~(\ref{eq:kappa3bound}), together with the perturbative-unitarity bounds from $hh\to hh$, allows signal strengths as large as $\mu_{3h}^{\text{HL-LHC}}\lesssim 100$. Imposing instead the projected constraint from~Eq.~(\ref{eq:kappa3boundfuture}) reduces this upper limit to $\mu_{3h}^{\text{HL-LHC}}\lesssim 85$. If, instead, $\kappa_3$ and $\kappa_{2t}$ are varied freely while fixing $\kappa_4=\kappa_t=1$ and $\kappa_{3t}=0$, the combination of Eq.~(\ref{eq:kappa3bound}) with Eq.~(\ref{eq:k2t_hh_LHCRunII}) gives $\mu_{3h}^{\text{HL-LHC}}\lesssim 130$. Using the projected constraint in Eq.~(\ref{eq:kappa3boundfuture}) reduces this upper limit to $\mu_{3h}^{\text{HL-LHC}}\lesssim 9$. Thus, scenarios in which only~$\kappa_3$ and $\kappa_{2t}$ depart from their SM values are unlikely to yield an observable triple-Higgs signal at the LHC. Indeed, imposing the projected constraint $0.77 < \mu_{2h}^{\text{HL-LHC}} < 1.23$~\cite{ATL-PHYS-PUB-2022-005} limits the maximal signal strength to $\mu_{3h}^{\text{HL-LHC}}\lesssim 25$ for $\kappa_4=\kappa_t=1$ and $\kappa_{3t}=0$. Similarly, allowing~$\kappa_4$ and~$\kappa_{2t}$ to vary freely, while fixing $\kappa_3=\kappa_t=1$ and $\kappa_{3t}=0$, and imposing~Eq.~(\ref{eq:k2t_hh_LHCRunII}), gives $\mu_{3h}^{\text{HL-LHC}}\lesssim 110$. By contrast, allowing either $\kappa_3$ or~$\kappa_4$ to vary together with $\kappa_{3t}$, subject to the perturbative-unitarity constraints from~$hh\to hh$ and $t\bar t\to3h$, respectively, while fixing all remaining $\kappa$ parameters to their SM values, can readily yield signal strengths exceeding the current LHC~Run~2 limit, $\mu_{3h}^{\text{LHC~Run~2}} < 588$~\cite{CMS:2026xkc}, even after imposing the projected constraint from~Eq.~(\ref{eq:kappa3boundfuture}) where applicable. This striking enhancement reflects the pronounced dependence of the $gg\to3h$ cross section on~$\kappa_{3t}$. Since~$\kappa_{3t}$ is, at tree level, unconstrained by measurements other than triple-Higgs production, it plays a particularly distinctive role in~$gg\to3h$ and motivates dedicated benchmark scenarios involving~$\kappa_{3t}$.

\subsection{\texorpdfstring{\textbf{\boldmath{Differential distributions}}}{Differential distributions}}
\label{sec:differential}

\begin{figure}[t!]
\begin{center}
\includegraphics[width=0.45\textwidth]{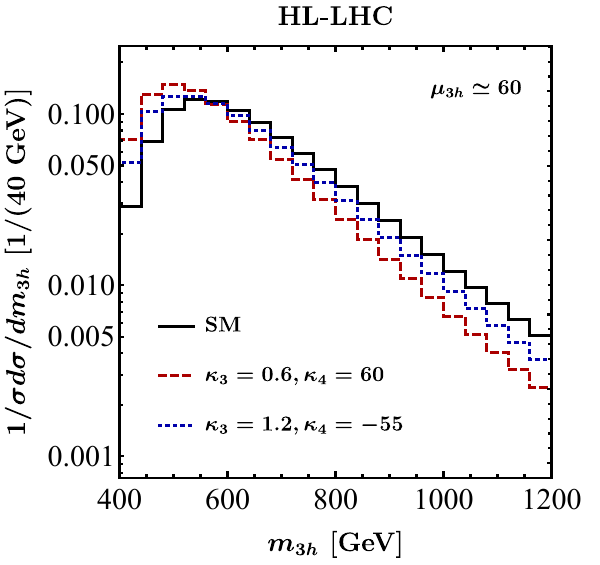} \quad 
\includegraphics[width=0.45\textwidth]{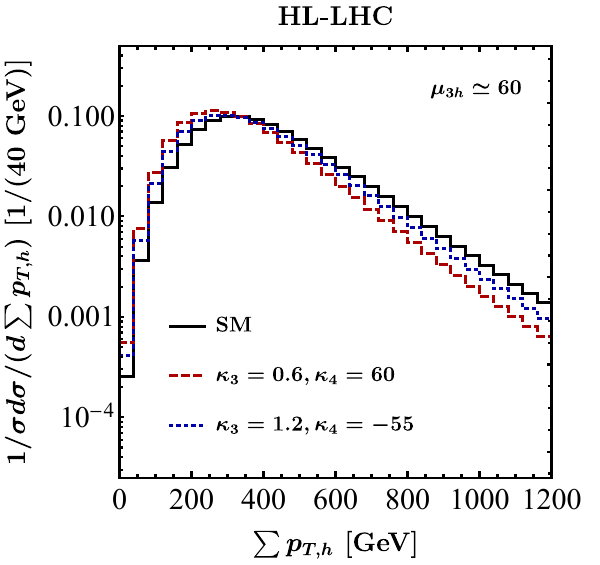}

\hspace{0mm}

\includegraphics[width=0.45\textwidth]{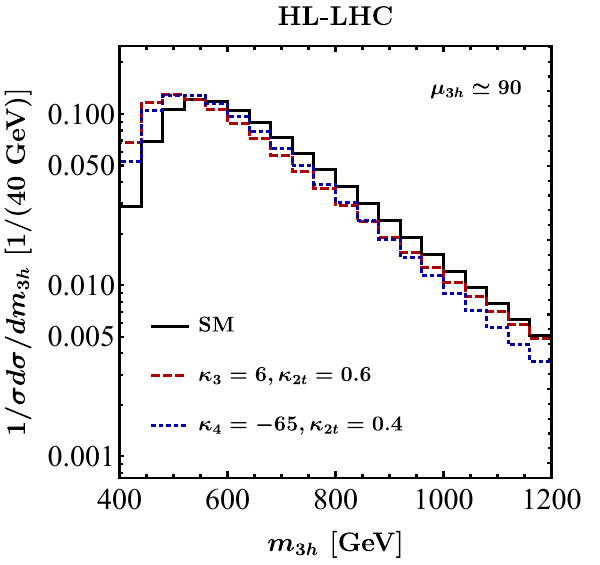} \quad 
\includegraphics[width=0.45\textwidth]{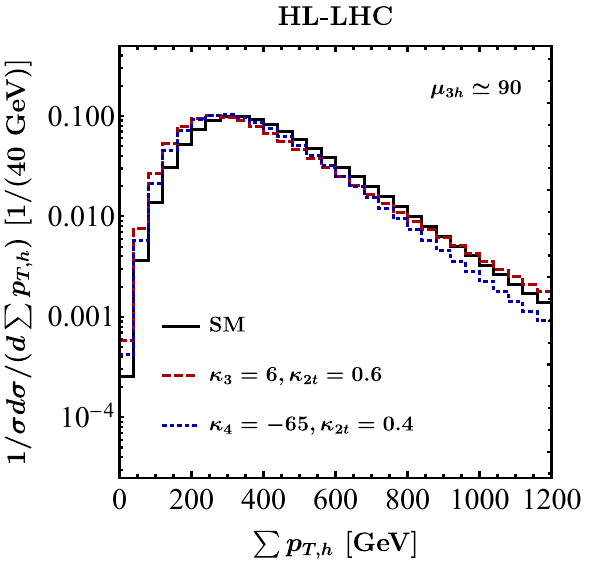}

\end{center}
\vspace{-2mm} 
\caption{\label{fig:distributions} Normalised kinematic distributions for $gg\to3h$ production at the HL-LHC in the SM (solid black line) and for four $\kappa$-benchmark scenarios indicated in the panels (red dashed and blue dotted lines). The $\kappa$ parameters not indicated in the panels are fixed to their SM values, as specified below~Eq.~(\ref{eq:Lkappa}). The left and right panels show the invariant mass of the three-Higgs system, $m_{3h}$, and the scalar sum of the Higgs-boson transverse momenta, $\textstyle\sum p_{T,h}$, respectively. The BSM predictions in the upper and lower panels correspond to $\mu_{3h}^{\mathrm{HL-LHC}}\simeq60$ and $\mu_{3h}^{\mathrm{HL-LHC}}\simeq90$, respectively. Further details are given in the text.}
\end{figure}

\begin{figure}[t!]
\begin{center}
\includegraphics[width=0.45\textwidth]{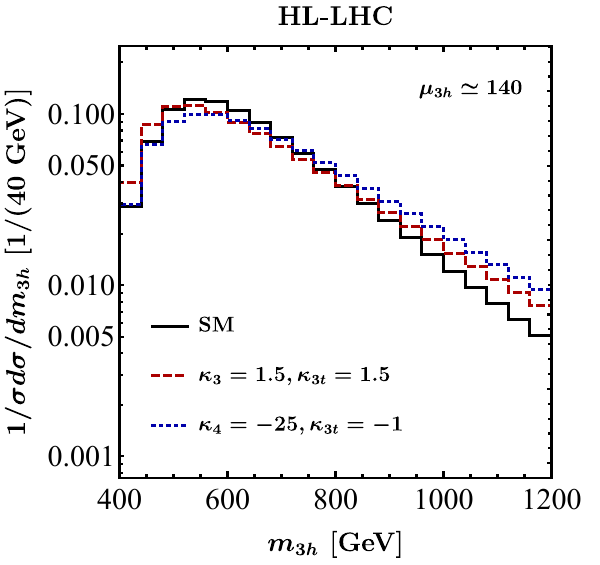} \quad 
\includegraphics[width=0.45\textwidth]{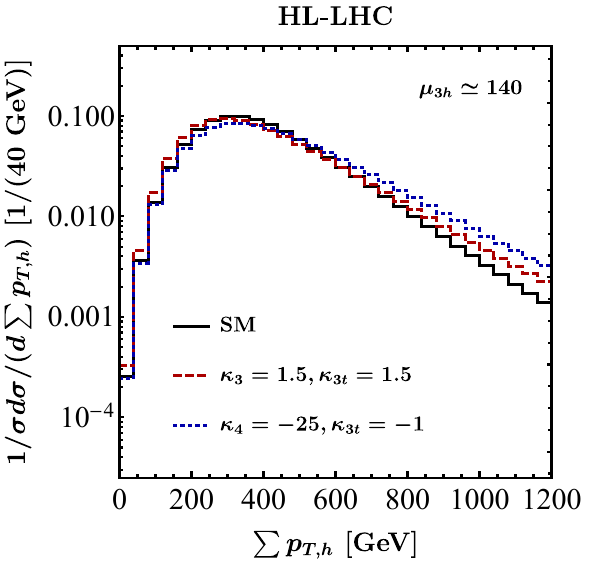}

\hspace{0mm}

\includegraphics[width=0.45\textwidth]{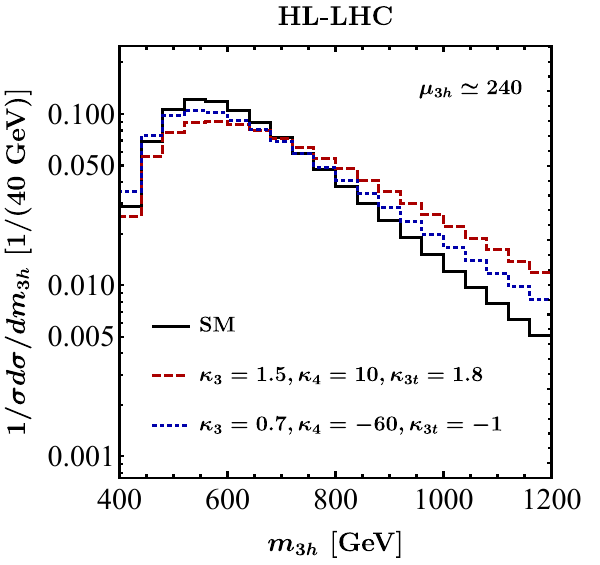} \quad 
\includegraphics[width=0.45\textwidth]{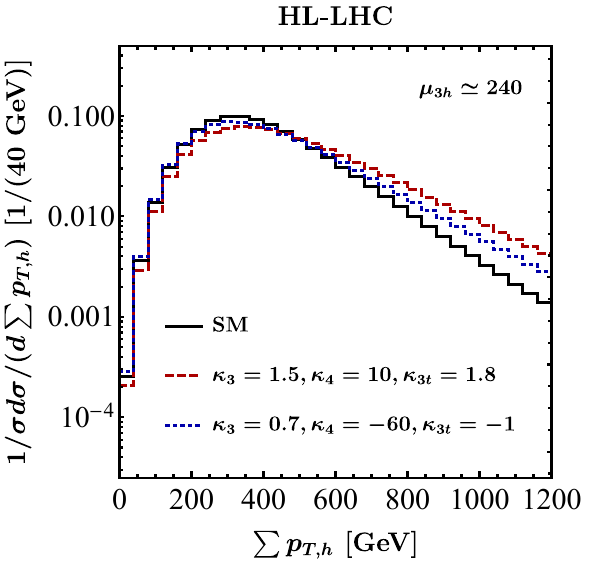}

\end{center}
\vspace{-2mm} 
\caption{\label{fig:distributions2} Same as~Figure~\ref{fig:distributions}, but for four additional benchmark scenarios in the~$\kappa$-framework, as indicated in the panels. The $\kappa$ parameters not indicated in the panels are fixed to their SM values, as specified below~Eq.~(\ref{eq:Lkappa}). The BSM predictions in the upper and lower panels correspond to $\mu_{3h}^{\mathrm{HL-LHC}}\simeq140$ and $\mu_{3h}^{\mathrm{HL-LHC}}\simeq240$, respectively.}
\end{figure}

Modifications of the trilinear and quartic Higgs self-couplings affect not only the inclusive triple-Higgs production rate but also the kinematic properties of the $gg\to3h$ final state. Figures~\ref{fig:distributions} and~\ref{fig:distributions2} illustrate this behaviour by showing normalised distributions at the HL-LHC for the SM and eight BSM benchmark scenarios. The $\kappa$ parameters not indicated in the legends are fixed to their SM values, as specified below~Eq.~(\ref{eq:Lkappa}). The left and right panels show the invariant mass of the three-Higgs system,~$m_{3h}$, and the scalar sum of the Higgs transverse momenta,~$\textstyle\sum p_{T,h}$, respectively. All benchmark points satisfy the perturbative-unitarity constraints discussed above and are arranged in pairs with nearly identical inclusive rates. The upper and lower rows in~Figure~\ref{fig:distributions} correspond to $\mu_{3h}^{\mathrm{HL-LHC}}\simeq60$ and $\mu_{3h}^{\mathrm{HL-LHC}}\simeq90$, respectively, while those in~Figure~\ref{fig:distributions2} correspond to $\mu_{3h}^{\mathrm{HL-LHC}}\simeq140$ and $\mu_{3h}^{\mathrm{HL-LHC}}\simeq240$. Within each pair, the inclusive cross sections differ by less than about $5\%$.

The upper-row benchmarks in~Figure~\ref{fig:distributions} differ only in~$\kappa_3$ and~$\kappa_4$. Despite their nearly identical inclusive rates, their $m_{3h}$ and $\textstyle\sum p_{T,h}$ distributions differ by up to about~$50\%$, particularly around the peaks and in the high-energy tails. Here, we set $\kappa_t=1$ and $\kappa_{2t}=\kappa_{3t}=0$. Variations of~$\kappa_t$ predominantly rescale the unnormalised spectra and therefore have little impact on their normalised shapes~\cite{Panizzi:2025sya}. Differential information can thus distinguish benchmark scenarios that are nearly degenerate at the level of the inclusive rate, potentially extending the HL-LHC sensitivity beyond the projected interval in~Eq.~(\ref{eq:kappa4boundsHLLHC}), which is based on the inclusive rate alone~\cite{Ding:2026qto}. The lower-row benchmarks in~Figure~\ref{fig:distributions} exhibit a similar degeneracy, with either~$\kappa_3$ or~$\kappa_4$ accompanied by a non-SM value of~$\kappa_{2t}$. Although their inclusive rates are nearly identical, the corresponding kinematic shapes differ, illustrating how differential observables can help disentangle different combinations of Higgs self-couplings and top-Higgs contact interactions. We note, however, that all benchmarks shown in~Figure~\ref{fig:distributions} lie below the projected HL-LHC upper limit on the signal strength. Exploiting their shape differences will therefore require improved experimental~sensitivity.

The benchmarks in~Figure~\ref{fig:distributions2} illustrate the impact of~$\kappa_{3t}$ on the kinematic properties of triple-Higgs production. Since their signal strengths exceed the projected HL-LHC upper limit, $\mu_{3h}^{\text{HL-LHC}} < 125$, their distinct kinematic shapes are particularly relevant for future searches and could provide additional sensitivity to scenarios with non-zero~$\kappa_{3t}$. BSM scenarios with non-vanishing~$\kappa_{3t}$ and non-SM values of either~$\kappa_3$ or~$\kappa_4$ can yield signal strengths above the projected HL-LHC sensitivity, as shown in the upper panels. The most pronounced shape differences arise, however, for parameter choices in which~$\kappa_3$,~$\kappa_4$, and~$\kappa_{3t}$ all deviate from their SM values, as illustrated in the lower panels. This behaviour arises because, in this case, the interference between the different topologies contributing to~$gg \to 3h$ is enhanced. Taken~together, Figures~\ref{fig:distributions} and~\ref{fig:distributions2} demonstrate that differential information can resolve parameter degeneracies that are invisible in the inclusive rate and provide additional sensitivity to both Higgs self-couplings and top-Higgs contact interactions. This motivates the use of kinematic observables alongside the inclusive rate in future HL-LHC searches for $gg\to3h$ production.

\section{\textbf{\boldmath{Conclusions}}}
\label{sec:conclusions}

In contrast to the Higgs couplings to EW gauge bosons and third-generation fermions, which are already being probed with increasing precision, the Higgs self-couplings remain largely unconstrained by LHC~Run~2 data. Considerable room therefore remains for BSM scenarios with substantial modifications of the Higgs potential. Indeed, both theoretical considerations and explicit ultraviolet-complete models show that deviations of the Higgs self-couplings exceeding $200\%$ can remain compatible with current single-Higgs measurements without requiring fine-tuned parameters~\cite{Durieux:2022hbu}. Measurements of multi-Higgs production at the HL-LHC and future colliders therefore offer a unique opportunity to explore largely unconstrained regions of parameter space, with significant discovery potential even when single-Higgs observables remain close to their SM predictions. A similar situation arises for top-Higgs contact interactions, which can also receive sizable modifications in BSM scenarios with a non-linear Higgs realisation~\cite{Gillioz:2012se}.

In this note, we have derived compact ``pocket formulae'' for gluon-fusion triple-Higgs production within the~$\kappa$-framework, valid at LO in perturbation theory and retaining the full dependence on~$\kappa_3$,~$\kappa_4$,~$\kappa_t$,~$\kappa_{2t}$, and~$\kappa_{3t}$ at COM energies of $\sqrt{s}=13\,\mathrm{TeV}$, $\sqrt{s}=13.6\,\mathrm{TeV}$, and $\sqrt{s}=14\,\mathrm{TeV}$. Since our derivation is based on flexible matrix-element implementations, extending the ``pocket formulae'' to additional coupling modifications, such as those arising at next-to-leading order in the HEFT~\cite{Brivio:2025sib}, is comparatively straightforward. The ``pocket formulae'' are provided in machine-readable form as ancillary files with the arXiv submission of this note, at full numerical precision and together with the MC uncertainties of all coefficients and their correlations. Our results complement those of the LHC Higgs Working Group study in~Ref.~\cite{Bethani:2026}, which focuses on modelling inclusive and differential distributions in the~HEFT and also covers $t\bar{t}hh$ production. Taken together, the two works provide a consistent basis for interpreting triple-Higgs searches at the LHC, and we encourage readers to use them side~by~side.

Applying these parametrisations to current and projected LHC data, we find that the present uncertainty in~$\kappa_t$ has a visible impact on the resulting constraints in the~$\kappa_3\hspace{0.25mm}$--$\hspace{0.25mm}\kappa_4$ plane. This~effect was also noted in~Ref.~\cite{Panizzi:2025sya}. This dependence is reduced at the HL-LHC, where the top-quark Yukawa coupling is expected to be determined with a precision of a few percent. For~$\kappa_{2t}$, which parametrises the two-top-two-Higgs contact interaction, triple-Higgs production is unlikely to provide stronger constraints than double-Higgs production in a single-parameter analysis. It nevertheless provides complementary information in global fits, particularly once~$\kappa_4$ is included, as this coupling remains only weakly constrained by double-Higgs~\cite{Bizon:2018syu,Borowka:2018pxx,Li:2024iio,Bizon:2024juq,Ding:2026qto} and quadruple-Higgs~\cite{Papaefstathiou:2026sfv} production. For~$\kappa_{3t}$, which parametrises the two-top-three-Higgs contact interaction, we obtain the first constraint on this coupling from existing LHC data, as given in~Eq.~(\ref{eq:k3t_hhh_LHCRunII}). Its pronounced impact on triple-Higgs production, already for $\mathcal{O}(1)$ values, arises from the contact-interaction topology shown as the last diagram in~Figure~\ref{fig:diagrams}. Since this contribution does not involve an internal Higgs propagator, it is not subject to a propagator suppression. Gluon-fusion triple-Higgs production therefore offers a particularly sensitive~probe~of~$\kappa_{3t}$.

Our results further demonstrate that parameter choices leading to nearly identical inclusive $gg\to3h$ rates can nevertheless produce distinct kinematic distributions. This is true both for different combinations of~$\kappa_3$ and~$\kappa_4$ and for scenarios involving the top-Higgs contact interactions parametrised by~$\kappa_{2t}$ and~$\kappa_{3t}$. Particularly interesting are BSM scenarios with non-zero~$\kappa_{3t}$ and non-SM values of~$\kappa_3$ and/or~$\kappa_4$, which can simultaneously yield signal strengths above the projected HL-LHC upper limit, $\mu_{3h}^{\text{HL-LHC}} < 125$~\cite{ATL-PHYS-PUB-2025-003}, and distinctive kinematic shapes. We have identified, in particular, mixed $\kappa_3$, $\kappa_4$, and $\kappa_{3t}$ benchmark scenarios with nearly identical inclusive rates, $\mu_{3h}^{\text{HL-LHC}}\simeq240$, but visibly different normalised kinematic distributions, both from the SM prediction and from one another. These examples illustrate how differential information can resolve parameter degeneracies that remain hidden in the inclusive rate and thereby provide additional sensitivity to both Higgs self-couplings and top-Higgs contact interactions in future triple-Higgs searches.

The projected HL-LHC sensitivity to~$\kappa_4$ in a single-parameter analysis remains modest, with constraints reaching only $|\kappa_4|\lesssim 80$, comparable to those derived from perturbative-unitarity arguments~\cite{Liu:2018peg,Stylianou:2023tgg,DiLuzio:2017tfn}. The proposed hadron-collider stage of the Future Circular Collider could improve the sensitivity to the quartic Higgs self-coupling by approximately one order of magnitude~\cite{Papaefstathiou:2015paa,Chen:2015gva,Fuks:2015hna,Kilian:2017nio,Fuks:2017zkg,Bizon:2018syu,Papaefstathiou:2023uum,Abouabid:2024gms,Bizon:2024juq,Dong:2025lkm,Haisch:2025pql,Fuks:2025gjv}. Achieving a precision at the level of $50\%$, however, would likely require the capabilities of a future muon collider~\cite{Chiesa:2020awd}.

\section*{Acknowledgements}

This work was carried out within the LHC Higgs Working Group as a contribution to the CERN Report~5. We thank the members of the working group for valuable discussions, and Alexandra~Carvalho, Ennio Salvioni, Angela Taliercio, and Eleni Vryonidou for their helpful comments, which allowed us to improve the manuscript. UH thanks Konstantin Schmid for helpful discussions. We acknowledge the use of Claude Opus~5.5 (Anthropic), an artificial-intelligence assistant, which cross-checked the man-made \texttt{MCFM} implementation of the $gg \to 3h$ amplitude with anomalous Higgs couplings and helped improve the accuracy of the parametrisations in Eqs.~(\ref{eq:c00_13})--(\ref{eq:c02_14}) by deriving them from an exact decomposition of the squared amplitude instead of a~fit. All results were checked by the authors, who take full responsibility for the content of this~work. This~manuscript has been authored by Fermi Forward Discovery Group, LLC under Contract~No. 89243024CSC000002 with the U.S.~Department of Energy, Office of Science, Office of High Energy Physics. GDL~acknowledges support from the European Union's Horizon Europe research and innovation programme under the Marie Skłodowska-Curie grant agreement No. 101270050. RKE~acknowledges receipt of a Leverhulme Emeritus Fellowship from the Leverhulme Trust. AP acknowledges support from the National Science Foundation under Grant No.~PHY 2210161 and from the U.S.~Department of Energy, Office of Science, Office of Nuclear Physics, under Award No.~DE-SC0025728.

\begin{appendix}

\section{Perturbative-unitarity bounds on top-Higgs contact interactions}
\label{app:unitarity}

The top-Higgs contact interactions proportional to the coupling modifiers~$\kappa_{2t}$ and~$\kappa_{3t}$, introduced in Eq.~(\ref{eq:Lkappa}), generate tree-level amplitudes for $t\bar t\to hh$ and $t\bar t\to 3h$. In the high-energy limit $\hat s\gg m_t^2$, where $\hat s$ denotes the squared partonic COM energy of the $t\bar t$ system, the scalar bilinear $\bar t t$ selects equal physical helicities for the top and antitop. Equivalently, the spinor product $\bar v(p_2,\lambda_2)\,u(p_1,\lambda_1)$ vanishes for $\lambda_1\neq\lambda_2$, up to corrections suppressed by~$m_t/\sqrt{\hat s}$. Thus,~at leading order, only the two helicity configurations $(\lambda_1,\lambda_2)=(\pm1/2,\pm1/2)$ contribute. In~the standard convention, the physical antitop helicity is $\lambda_2$, while its spin projection along the~$z$-axis is $-\lambda_2$, since the antitop moves in the $-z$ direction. Consequently, the total angular-momentum projection is $J_z=\lambda_1-\lambda_2=0$. The resulting amplitudes are therefore dominated by the $J=0$ partial wave.

In the high-energy limit, averaging over the initial top spins and accounting for the identical Higgs bosons in the final state, the total cross sections are given by
\beq \label{eq:largesXs}
\sigma \left (t\bar t\to hh \right ) = \frac{\kappa_{2t}^2\hspace{0.25mm}y_t^2}{32\hspace{0.125mm}\pi} \, \frac{1}{v^2} \,, \qquad \sigma \left (t\bar t\to 3h \right ) = \frac{3\hspace{0.125mm}\kappa_{3t}^2\hspace{0.25mm}y_t^2}{1024\hspace{0.125mm}\pi^3} \, \frac{\hat s}{v^4} \,,
\eeq
where the two expressions make use of the integrated massless two- and three-body phase-space volumes $\Phi_2(\hat s)=1/(8\pi)$ and $\Phi_3(\hat s)=\hat s/(256\hspace{0.125mm}\pi^3)$, respectively, together with the $2!$ and~$3!$ symmetry factors for identical final-state Higgs bosons. Since the top-Higgs contact interactions are colour neutral, only the colour-singlet $t\bar t$ channel contributes. The cross sections above are written for this channel. Colour-averaging would introduce an overall factor of $1/3$, but the perturbative-unitarity bound is rescaled by the same factor, so the resulting bounds on~$\kappa_{2t}$ and~$\kappa_{3t}$ are unchanged.

For a two-particle initial state in a definite helicity configuration, perturbative unitarity bounds the exclusive cross section associated with a single partial wave $J$ according to
\beq \label{eq:sigmaexcluni}
\sigma_{\rm excl}^J \le \frac{4\pi \left ( 2 \hspace{0.125mm} J+1 \right )}{\hat s} \,.
\eeq
Here $\sigma_{\rm excl}^J$ refers to a fixed initial helicity configuration. Since only two of the four initial spin combinations give non-vanishing amplitudes, averaging over the initial spins gives $\sigma = \sigma_{\rm excl}^J/2$. Saturating the $J=0$ partial wave then gives
\beq \label{eq:sigma0uni}
\sigma \le \frac{2\pi}{\hat s} \,.
\eeq
This is consistent with Bose symmetry in the $t\bar t\to hh$ process. Since the contact amplitude is isotropic, it is trivially invariant under the exchange of the two identical Higgs bosons, which corresponds to the transformation $\theta\to\pi-\theta$. The relevant Wigner matrix element is $d^J_{0,0}(\theta)=P_J(\cos\theta)$, where $P_J(\cos\theta)$ denotes the $J^{\rm th}$ Legendre polynomial, and it satisfies $d^J_{0,0}(\pi-\theta)=(-1)^J d^J_{0,0}(\theta)$. Therefore, Bose symmetry requires the partial-wave expansion to contain only even values of $J$, with $J=0$ being the lowest allowed partial wave.

Applying the bound~(\ref{eq:sigma0uni}) to the two cross sections in~Eq.~(\ref{eq:largesXs}) yields
\beq \label{eq:boundsanal}
|\kappa_{2t}| \le 8\hspace{0.25mm}\pi\hspace{0.25mm}\frac{v}{y_t\sqrt{\hat s}} \,, \qquad |\kappa_{3t}| \le 32\hspace{0.25mm}\sqrt{\frac{2}{3}}\hspace{0.5mm}\pi^2\hspace{0.25mm}\frac{v^2}{y_t \hspace{0.25mm} \hat s} \,,
\eeq
in agreement with the dimensional-analysis expectation that the modifier~$\kappa_{2t}$ associated with the dimension-five interaction is bounded as $1/\sqrt{\hat s}$, while the modifier~$\kappa_{3t}$ associated with the dimension-six interaction is bounded as $1/\hat s$. These bounds apply to the physical Higgs boson~only. The corresponding bounds obtained from the full EW-symmetric dimension-six operator in the SM effective field theory, which also includes the would-be Goldstone-boson channels, are necessarily stronger, as demonstrated explicitly, for example, in~Ref.~\cite{Bresciani:2025toe}.

To ensure that our predictions for the high-energy tails of the~$gg\to 3h$ differential distributions remain within the regime of validity of the~$\kappa$-framework, we restrict our analysis to partonic COM energies below the scale at which the perturbative unitarity bounds on the $t\bar t\to hh$ and $t\bar t\to 3h$ subamplitudes are saturated. Guided by the distributions shown in~Figures~\ref{fig:distributions} and~\ref{fig:distributions2}, we adopt $\sqrt{\hat s}=2\,{\rm TeV}$ as a representative benchmark. At this energy and for~$y_t \simeq 0.94$, we~obtain
\beq \label{eq:boundsnume}
|\kappa_{2t}| \le 3.3 \,, \qquad |\kappa_{3t}| \le 4.2 \,.
\eeq
These bounds are used in the analysis presented in~Section~\ref{sec:results}.

\end{appendix}



\end{document}